\documentclass[a4paper,twocolumn,noshowpacs,preprintnumbers]{revtex4}

\usepackage[utf8x]{inputenc}
\usepackage{epsfig,amssymb,amsmath,psfrag}

\usepackage{etex}
\usepackage{pstricks-add}

\usepackage{bm}
\usepackage{epic}
\usepackage{graphicx}
\usepackage{color}
\usepackage{hyperref}
\usepackage{young}
\usepackage{multirow}
\usepackage{fancybox}

\usepackage[paperwidth=210mm,paperheight=297mm,centering,hmargin=1.75cm,vmargin=2.5cm]{geometry}

\def \be  {\begin{equation}}
\def \ee  {\end{equation}}
\def \ba  {\begin{eqnarray}}
\def \ea  {\end{eqnarray}}
\def \baa {\begin{eqnarray*}}
\def \eaa {\end{eqnarray*}}
\def \bb  {\begin {thebibliography} }
\def \eb  {\end{thebibliography}}
\def \lab #1 {\label{#1}}

\newcommand{\nn}{\nonumber}
\newcommand{\beq}{\begin{equation}}
\newcommand{\eeq}{\end{equation}}
\newcommand{\beqa}{\begin{eqnarray}}
\newcommand{\eeqa}{\end{eqnarray}}

\newcommand{\tlambda}{\tilde{\lambda}}

\newcommand{\NN}{\mathcal{N}}

\newcommand{\sfrac}[2]{{\textstyle\frac{#1}{#2}}}
\newcommand{\half}{\sfrac{1}{2}}

\begin{document}

\preprint{HU-EP-12/50,
HU-Mathematik:14-2012, DESY 12-228, ZMP-HH/12-26, AEI-2012-198}

\title{Harmonic R-matrices for Scattering Amplitudes and Spectral Regularization}
\author{Livia Ferro${}^{1}$}
\author{Tomasz \L ukowski${}^{2}$}
\author{Carlo Meneghelli${}^{3}$}
\author{Jan Plefka${}^1$}
\author{Matthias Staudacher${}^{2,4}$}
\affiliation{
${}^1${
Institut f\"ur Physik,
Humboldt-Universit\"at zu Berlin,
Newtonstra{\ss}e 15, 12489 Berlin, Germany}\\
${}^2${
Institut f\"ur Mathematik und Institut f\"ur Physik, Humboldt-Universit\"at zu Berlin,\\
IRIS Adlershof, 
Zum Gro\ss en Windkanal 6, 12489 Berlin, Germany}\\
${}^3${
Fachbereich Mathematik, Universit\"at Hamburg, Bundesstra\ss e 55, 20146 Hamburg,\\
\& Theory Group, DESY, Notkestra\ss e 85, 22603 Hamburg, Germany}\\
${}^4${
MPI f\"ur Gravitationsphysik, Albert-Einstein-Institut,
Am M\"uhlenberg 1, 14476 Potsdam, Germany
}\\
{\tt ferro,plefka$\bullet$physik.hu-berlin.de, carlo.meneghelli$\bullet$gmail.com, lukowski,staudacher$\bullet$mathematik.hu-berlin.de}}
\begin{abstract}

Planar $\mathcal{N}=4$ super Yang-Mills appears to be integrable. While this allows to find this theory's exact spectrum, integrability has hitherto been of no direct use for scattering amplitudes. To remedy this, we deform all scattering amplitudes by a spectral parameter. The deformed tree-level four-point function turns out to be essentially the one-loop R-matrix of the integrable $\mathcal{N}=4$ spin chain satisfying the Yang-Baxter equation. Deformed on-shell three-point functions yield novel three-leg R-matrices satisfying bootstrap equations. Finally, we supply initial evidence that the spectral parameter might find its use as a novel symmetry-respecting regulator replacing dimensional regularization. Its physical meaning is a local deformation of particle helicity, a fact which might be useful for a much larger class of non-integrable four-dimensional field theories.

\end{abstract}

\maketitle

\section{Introduction}

Amazing features have been discovered in the last years in studying the structure of planar 
maximally supersymmetric Yang-Mills theory ($\NN = 4$ SYM).
The discovery of a hidden dual superconformal symmetry \cite{Drummond:2008vq}, after combining with the conventional superconformal symmetry into a Yangian structure \cite{Drummond:2009fd}, points  to an underlying integrability.
This structure is deeply connected to the Gra\ss mannian formulation of scattering amplitudes \cite{ArkaniHamed:2009dn,Mason:2009qx}.  Here the tree-level $n$-point N$^{k-2}$MHV amplitudes can be written as
\begin{equation}
\label{gras.amp}
\mathcal{A}_{n,k}^{\mathrm{tree}}=\oint\frac{\prod_{a=1}^{k}\prod_{i=k+1}^n dc_{ai}}{\mathcal{M}_1\mathcal{M}_2\ldots \mathcal{M}_n}\delta^{4|4}\left(C_{(k,n)}\cdot \mathcal{Z}\right)\,,
\end{equation}
where $\mathcal{Z}_i^{\mathcal{A}}$ are the super-twistor variables 
$(\tilde \mu^{\alpha}_{i},{\tilde \lambda}^{\dot\alpha}_{i}, \eta^{A}_{i})$ with $\tilde \mu^{\alpha}_{i}$
 the Fourier conjugate to $\lambda_{i}^{\alpha}$, and 
 $\mathcal{A}$ is a fundamental index of $\mathfrak{gl}(4|4)$.  Recall that the momenta of scattering amplitudes are expressed as $p_i^{\alpha\dot\alpha} = \lambda_i^{\alpha} \tilde\lambda_i^{\dot\alpha}$, and $\eta_i^A$ are Gra\ss mann variables. 
Moreover, $C_{(k,n)}$ stands for a $(k\times n)$ matrix of the complex parameters $c_{ai}$, and the first $k$ columns have been fixed to a unit matrix using the GL$(k)$ symmetry of the integral. By $\mathcal{M}_i$ we denote the $(k\times k)$ minors of the $C_{(k,n)}$ matrix. 
In a remarkable, very recent construction \cite{ArkaniHamed} all amplitudes are argued to be constructible to arbitrary loop order in terms of basic on-shell building blocks through BCFW recursion relations \cite{Britto:2004ap}. 
More precisely, any amplitude at arbitrary but fixed loop order is expressible as a sum over suitable on-shell diagrams obtained by appropriately linking $\mathrm{MHV}$ and $\overline{\mathrm{MHV}}$ three-point amplitudes and subsequently  integrating out all on-shell super-twistor variables on internal links.

In a seemingly unrelated recent development, a connection between tree-level amplitudes and the complete one-loop dilatation operator was pointed out in \cite{Zwiebel:2011bx}. In particular the Hamiltonian of the $\mathcal{N}=4$ spin chain was shown to be related to the tree-level four-point amplitude. Being {\it integrable}, this nearest-neighbor Hamiltonian is generated by an R-matrix satisfying the celebrated Yang-Baxter equation \cite{Beisert:2003yb}. After defining monodromy matrices, R-matrices serve as an alternative, and from the perspective of scattering processes more natural, way to define the Yangian algebra. The crucial feature of R-matrices is their dependence on a complex parameter called spectral parameter. So far, the fundamental  question on how to insert the spectral parameter into the scattering amplitude problem had not yet been asked, let alone answered. In this letter we fill this gap by first unifying and generalizing the mentioned developments. We then proceed to the investigation of radiative corrections to scattering amplitudes. Excitingly, we find preliminary one-loop evidence that the introduction of appropriate spectral parameters allows to regulate all infrared divergences while staying in strictly four dimensions, and more generally locally respecting all symmetries.

The structure is as follows. In \ref{sec.Rmatrix}.~we start from the Yang-Baxter equation and find its solution in terms of a spectral-parameter dependent deformation of the four-point tree-level scattering amplitude. In \ref{sec.blocks}.~we then construct the deformed three-point building blocks of this R-matrix and relate the spectral parameter to the central charge of particles involved in the scattering process, which in turn leads to a physical interpretation of the deformation as a relaxation of the helicity constraints on particles. In \ref{sec.specreg}.~we present our proposal for the spectral regularization of loop amplitudes. Section \ref{sec.outlook}.~provides conclusions and an outlook.

\section{Gra\ss mannian R-matrix}
\label{sec.Rmatrix}

As the first step in our construction we find a spectral parameter
dependent deformation of the tree-level MHV four-point amplitude. It is given by an
R-matrix which can be found from the Yang-Baxter equation
written in the tensor product of two super-twistor spaces, labeled 1 and 2, and the
fundamental space, labeled 3
\begin{equation*}
\mathbf{R}_{12}(z_3) \mathbf{R}_{13}(z_2) \mathbf{R}_{23}(z_1) =
\mathbf{R}_{23}(z_1) \mathbf{R}_{13}(z_2) \mathbf{R}_{12}(z_3)\,,
\end{equation*}
where $z_1$, $z_2$, and $z_3=z_2-z_1$ are spectral parameters.
The well-known R-matrices acting on the tensor product of the fundamental and super-twistor spaces are given by 
\begin{equation*}
\mathbf{R}_{i3,\mathcal{B}}^{\,\mathcal{A}}(z) = z\,
\delta_{\mathcal{B}}^{\mathcal{A}} + (-1)^{\mathcal{B}} J_{i\,
\mathcal{B}}^{\mathcal{A}}\,,
\end{equation*}
where $ J_{i\,
\mathcal{B}}^{\mathcal{A}}=\mathcal{Z}_{i}^{\mathcal{A}}\frac{\partial}{\partial\mathcal{Z}_i^{\mathcal{B}}}$
are the generators of a twistor rep of $\mathfrak{gl}(4|4)$ and $ (-1)^{\mathcal{A}} $ encodes grading.
Then the Yang-Baxter equation is a linear equation for the R-matrix
$\mathbf{R}_{12}(z)$ intertwining two super-twistor representations.

Let us call $\mathcal{R}(z)$ the integral kernel of
$\mathbf{R}_{12}(z)$. We look for a solution of the Yang-Baxter equation
in Gra\ss mannian form, namely
\begin{equation*}
\mathcal{R}(z)=\oint\frac{dc_{13}dc_{14}dc_{23}dc_{24}}{c_{13}c_{24}(c_{13}
c_{24}-c_{14}c_{23})}{ F(C;z)}\delta^{4|4}(C_{(2,4)}\cdot \mathcal{Z})\,,
\end{equation*}
where we introduced the function $F(C;z)$. This function is uniquely determined by the Yang-Baxter equation together with the requirement that  all particles have physical helicities. One finds
\begin{equation*}
F(C;z)=\left(\frac{c_{13}c_{24}}{c_{13} c_{24}-c_{14} c_{23}}\right)^z \,.
\end{equation*}
Hereafter, we will refer to $\mathcal{R}(z)$ as the four-point harmonic R-matrix.
After specifying the integration over the $c$-variables, this is essentially the kernel of the one-loop R-matrix of the $\mathcal{N}=4$ spin chain of \cite{Beisert:2003yb}!
Excitingly, for $z\neq 0$, $\mathcal{R}(z)$ can also be interpreted as a deformation of the $n=4$ and $k=2$ expression in (\ref{gras.amp}). Similar but more complicated deformations exist for any $n$ and $k$ as we will discuss in the following.

In this letter we focus on the superamplitudes of $\mathcal{N}=4$ SYM but
a similar calculation can be done for any representation of
$\mathfrak{gl}(n|m)$ that can be written in terms of one family of
oscillators -- the so-called generalized one-row reps (see e.g.~\cite{Frassek:2011aa}). The result applies to more general integrable chains and is related to the harmonic action of their Hamiltonians described in \cite{Beisert:2004ry}. We
defer the construction to a separate paper \cite{Ferro2012}.

\section{Three-point R-matrices}
\label{sec.blocks}

\begin{figure}[h]
 \psset{unit=0.8cm}
 \begin{pspicture}(5,1)
 \rput(0,0){\rnode{A1}{1}}
 \rput(0.5,0.6){\rnode{A2}{}}
 \rput(0.5,1.2){\rnode{A3}{\,\,\,\,1}}
 \rput(1,0){\rnode{A4}{ 2}}
 \rput(-0.2,0.43){\rnode{A5}{3\,}}
 \rput(1.2,0.27){\rnode{A6}{}}
 \ncline{A1}{A2}
 \ncline{A2}{A3}
 \ncline{A4}{A2}
 \ncline{A5}{A6}
 \rput(1.5,0.6){\rnode{a}{=}}
 \rput(2,0){\rnode{B1}{1}}
 \rput(2.5,0.6){\rnode{B2}{}}
 \rput(2.5,1.2){\rnode{B3}{\,\,\,\,1}}
 \rput(3,0){\rnode{B4}{2}}
 \rput(1.8,0.93){\rnode{B5}{}}
 \rput(3.2,0.77){\rnode{B6}{\,3}}
 \ncline{B1}{B2}
 \ncline{B2}{B3}
 \ncline{B4}{B2}
 \ncline{B5}{B6}
 \pscircle[fillstyle=solid,fillcolor=black](0.5,0.6){0.12}
 \pscircle[fillstyle=solid,fillcolor=black](2.5,0.6){0.12}
 \end{pspicture}
 \begin{pspicture}(3,1)
 \rput(0,1.2){\rnode{A1}{1 }}
 \rput(0.5,0.6){\rnode{A2}{}}
 \rput(0.5,0){\rnode{A3}{\,\,\,\,1}}
 \rput(1,1.2){\rnode{A4}{ 2}}
 \rput(-0.2,0.77){\rnode{A5}{3\,}}
 \rput(1.2,0.93){\rnode{A6}{}}
 \ncline{A1}{A2}
 \ncline{A2}{A3}
 \ncline{A4}{A2}
 \ncline{A5}{A6}
 \rput(1.5,0.6){\rnode{a}{=}}
 \rput(2,1.2){\rnode{B1}{1 }}
 \rput(2.5,0.6){\rnode{B2}{}}
 \rput(2.5,0){\rnode{B3}{\,\,\,\,1}}
 \rput(3,1.2){\rnode{B4}{ 2}}
 \rput(1.8,0.27){\rnode{B5}{}}
 \rput(3.2,0.43){\rnode{B6}{\,3}}
 \ncline{B1}{B2}
 \ncline{B2}{B3}
 \ncline{B4}{B2}
 \ncline{B5}{B6}
 \pscircle[fillstyle=solid,fillcolor=white](0.5,0.6){0.12}
 \pscircle[fillstyle=solid,fillcolor=white](2.5,0.6){0.12}
 \end{pspicture}
 \caption{Bootstrap equations for the three-point R-matrices.}
\label{fig.3point}
 \end{figure}
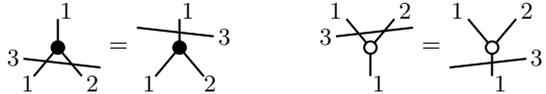

In very recent work it is demonstrated that the perturbative integrand of scattering amplitudes at arbitrary loop order naturally decomposes into basic cubic building blocks \cite{ArkaniHamed}. Encouragingly, this remains true under our deformation. In particular, one can find deformed three-point vertices which may subsequently be recombined into the R-matrix we found in the previous section.  As in the undeformed case \cite{ArkaniHamed}, there are two distinguished objects $\mathbf{R}_{\bullet}(z_1,z_2)$ and  $\mathbf{R}_\circ(z_1,z_2)$, which give deformations of the MHV and  $\overline{\mbox{MHV}}$ three-point
amplitudes, respectively. They satisfy the following bootstrap equations depicted in Fig.~\ref{fig.3point}, similar to but different from the Yang-Baxter equation of the previous section
\begin{eqnarray}\label{3point}\nonumber
\mathbf{R}_{\bullet}(z_1,z_2)
\mathbf{R}_{13}(0)\mathbf{R}_{23}(z_1)&=&z_1
\mathbf{R}_{13}(0)\mathbf{R}_{\bullet}(z_1,z_2)\,,\\ 
\mathbf{R}_{23}(z_1)\mathbf{R}_{13}(0)\mathbf{R}_{\circ}(z_1,z_2)&=&z_1
\mathbf{R}_{\circ}(z_1,z_2) \mathbf{R}_{13}(0)\,.\hspace{0.5cm}
\end{eqnarray}
An additional set of equations is obtained by replacing space 1 with space 2, leading to a second spectral parameter $z_2$.

 Once the
integral kernels $\mathcal{R}_{\bullet}(z_1,z_2)$ and
$\mathcal{R}_\circ(z_1,z_2)$ are defined one finds the following solutions to
\eqref{3point} in the Gra\ss mannian form
\begin{eqnarray}
\mathcal{R}_\bullet(z_1,z_2)&=&\oint \frac{dc_1 dc_2}{c_1 c_2}{
\frac{1}{c_1^{z_1} c_2^{z_2}}}\delta^{4|4}(C_{(2,3)}\cdot \mathcal{Z})\,,\nn \\
\mathcal{R}_\circ(z_1,z_2)&=&\oint \frac{dc_1 dc_2}{c_1
c_2}{\frac{1}{c_1^{z_1} c_2^{z_2}}}\delta^{4|4}(C_{(1,3)}\cdot \mathcal{Z})\, \nn \,.
\end{eqnarray}
After integration, the three-point R-matrices take, under the constraint $z_1+z_2+z_3=0$, a $\mathbb{Z}_3$-symmetric form strikingly similar to conformal field theory correlators 
\begin{eqnarray}
\label{Rz3}
\mathcal{R}_\bullet(z_1,z_2)&=&  \frac{\delta^{4}(p^{\alpha\dot\alpha}) \delta^{8}(q^{\alpha A}) }{\langle 1\,2\rangle^{1+z_3} \langle 2\,3\rangle^{1+z_1} \langle 3\,1\rangle^{1+z_2}} \,,\nn \\
\mathcal{R}_\circ(z_1,z_2)&=& \frac{\delta^{4}(p^{\alpha\dot\alpha}) \delta^{4}(\tilde q^A)}{ [1\,2]^{1+z_3} [2\,3]^{1+z_1} [3\,1]^{1+z_2}}  \,,
\end{eqnarray}
where we use the standard helicity spinor representations of momentum and
super-charges (see e.g. \cite{Drummond:2010ep}). 

\begin{figure}[h]
  \begin{pspicture}(1.7,2)
   \rput(0,0){\rnode{A1}{3}}
   \rput(0.4,0.4){\rnode{A2}{}}
   \rput(0.4,1.4){\rnode{A3}{}}
   \rput(0,1.8){\rnode{A4}{4}}
   \rput(1.4,1.4){\rnode{A5}{}}
   \rput(1.8,1.8){\rnode{A6}{1}}
   \rput(1.4,0.4){\rnode{A7}{}}
   \rput(1.8,0){\rnode{A8}{2}}
   \ncline{A1}{A2}
   \ncline{A2}{A3}
   \ncline{A3}{A4}
   \ncline{A3}{A5}
   \ncline{A5}{A6}
   \ncline{A5}{A7}
   \ncline{A7}{A8}
   \ncline{A2}{A7}
   \pscircle[fillstyle=solid,fillcolor=white](0.4,0.4){0.15}
   \pscircle[fillstyle=solid,fillcolor=black](1.4,0.4){0.15}
   \pscircle[fillstyle=solid,fillcolor=black](0.4,1.4){0.15}
   \pscircle[fillstyle=solid,fillcolor=white](1.4,1.4){0.15}
   \rput(0.9,0.9){\rnode{aaa}{z}}
   \psarc[arcsepB=1pt]{<-}(0.9,0.9){0.2}{50}{10}
   \end{pspicture}
   \caption{Four-point R-matrix from three-point R-matrices.}
    \label{fig.4pointR}
   \end{figure}
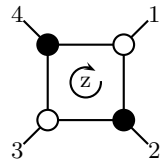

Again in generalization of an important insight of \cite{ArkaniHamed}, one
has to now glue four three-point R-matrices with appropriate spectral
parameters (see Fig.~\ref{fig.4pointR}) in order to reproduce the result for the four-point R-matrix of the last section. 
Exactly as in the undeformed case in \cite{ArkaniHamed}, it is important to stress that the R-matrix depicted in Fig.~\ref{fig.4pointR} is tree-level as opposed to one-loop.

It may be confusing that in our construction the three-point
R-matrices depend on two spectral parameters as opposed to the one parameter of the four-point
R-matrix. The reason is that for the latter
we additionally assumed that all external particles have physical helicities. It is
easy to check that when one makes this further assumption, solutions to \eqref{3point} cease to exist. In order to
obtain a non-trivial result one has to relax this condition. It is
then possible to find an interpretation of the spectral parameters by acting
with the central charges
$\mathcal{C}_i=\frac{1}{2}\sum_{\mathcal{A}}\mathcal{Z}_i^{\mathcal{A}}\frac{\partial}{\partial\mathcal{Z}_i^{\mathcal{A}}}$
\begin{eqnarray*}
\mathcal{C}_1\,
\mathcal{R}_{\circ}(z_1,z_2)&=&\half z_1\,\mathcal{R}_{\circ}(z_1,z_2)\,,\\
\mathcal{C}_2\,
\mathcal{R}_{\circ}(z_1,z_2)&=&\half z_2\,\mathcal{R}_{\circ}(z_1,z_2)\,,\\
\mathcal{R}_{\circ}(z_1,z_2)\,
\mathcal{C}_3&=&\half(z_1+z_2)\,\mathcal{R}_{\circ}(z_1,z_2)\,,
\end{eqnarray*}
and analogously for $\mathcal{R}_{\bullet}$. We see that the spectral
parameters have the interpretation of central charge eigenvalues of the three
particles, and that furthermore the vertices conserve the total central charge. 
Since the spectral parameter can be any complex number it means
that the particles carry non-zero central
charges, and accordingly unphysical helicities not restricted to integers or half-integers, as the (super-)helicity generator of the $i$th particle is $h_{i}=1-\mathcal{C}_{i}$. 

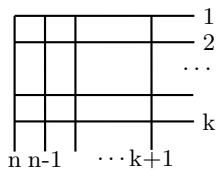
\begin{figure}[h]
\begin{pspicture}(3,2)
\rput(0.5,0.6){\rnode{A1}{}}
\rput(3,0.6){\rnode{A2}{ k}}
\rput(0.5,0.95){\rnode{B1}{}}
\rput(2.85,0.95){\rnode{B2}{}}
\rput(2.9,1.3){\rnode{C1}{$\ldots$}}
\rput(0.5,1.65){\rnode{D1}{}}
\rput(3,1.65){\rnode{D2}{ 2}}
\rput(0.5,2){\rnode{E1}{}}
\rput(3,2){\rnode{E2}{ 1}}
\ncline{A1}{A2}
\ncline{B1}{B2}
\ncline{D1}{D2}
\ncline{E1}{E2}
\rput(0.5,0.07){\rnode{F11}{n}}
\rput(0.5,0.2){\rnode{F1}{}}
\rput(0.5,2){\rnode{F2}{}}
\rput(0.9,0.1){\rnode{G1}{n-1}}
\rput(0.9,2){\rnode{G2}{}}
\rput(1.3,0.2){\rnode{H1}{}}
\rput(1.3,2){\rnode{H2}{}}
\rput(2.3,0.1){\rnode{I1}{k+1}}
\rput(2.3,2){\rnode{I2}{}}
\rput(1.8,0.1){\rnode{KK}{$\dots$}}
\ncline{F1}{F2}
\ncline{G1}{G2}
\ncline{H1}{H2}
\ncline{I1}{I2}
\end{pspicture}
\caption{Lattice encoding $\mathcal{R}_{n,k}$}
\label{fig.lattice}
\end{figure}

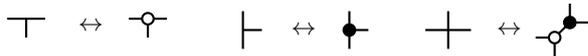
\begin{figure}[h]
\begin{pspicture}(7.4,0.8)
\rput(0,0.5){\rnode{a1}{}}
\rput(0.5,0.5){\rnode{a2}{}}
\rput(0.25,0.5){\rnode{a3}{}}
\rput(0.25,0.25){\rnode{a4}{}}
\ncline{a1}{a2}
\ncline{a3}{a4}
\rput(1.1,0.4){\rnode{a5}{$\leftrightarrow$}}
\rput(1.6,0.5){\rnode{b1}{}}
\rput(2.1,0.5){\rnode{b2}{}}
\rput(1.85,0.5){\rnode{b3}{}}
\rput(1.85,0.25){\rnode{b4}{}}
\ncline{b1}{b2}
\ncline{b3}{b4}
\put(1.85,0.5){\circle{0.15}}
\pscircle[fillstyle=solid,fillcolor=white](1.85,0.5){0.09}

\rput(3.1,0.6){\rnode{c1}{}}
\rput(3.1,0.1){\rnode{c2}{}}
\rput(3.1,0.35){\rnode{c3}{}}
\rput(3.35,0.35){\rnode{c4}{}}
\ncline{c1}{c2}
\ncline{c3}{c4}
\rput(3.9,0.35){\rnode{c5}{$\leftrightarrow$}}
\rput(4.5,0.6){\rnode{d1}{}}
\rput(4.5,0.1){\rnode{d2}{}}
\rput(4.5,0.35){\rnode{d3}{}}
\rput(4.75,0.35){\rnode{d4}{}}
\ncline{d1}{d2}
\ncline{d3}{d4}
\pscircle[fillstyle=solid,fillcolor=black](4.5,0.35){0.09}

\rput(5.8,0.6){\rnode{e1}{}}
\rput(5.8,0.1){\rnode{e2}{}}
\rput(5.5,0.35){\rnode{e3}{}}
\rput(6.1,0.35){\rnode{e4}{}}
\ncline{e1}{e2}
\ncline{e3}{e4}
\rput(6.6,0.35){\rnode{e5}{$\leftrightarrow$}}
\rput(7.4,0.7){\rnode{f1}{}}
\rput(7.4,0.45){\rnode{f2}{}}
\rput(7.65,0.45){\rnode{f3}{}}
\rput(7.2,0.25){\rnode{f4}{}}
\rput(7.2,0){\rnode{f5}{}}
\rput(6.95,0.25){\rnode{f6}{}}
\ncline{f1}{f2}
\ncline{f2}{f3}
\ncline{f2}{f4}
\ncline{f4}{f5}
\ncline{f4}{f6}
\pscircle[fillstyle=solid,fillcolor=black](7.4,0.45){0.09}
\pscircle[fillstyle=solid,fillcolor=white](7.2,0.25){0.09}
\end{pspicture}
\caption{Dictionary for plabic diagrams.}
\label{fig.dictionary}
\end{figure}

There exists a simple way to produce higher-point harmonic R-matrices
by gluing only three-point R-matrices. Taking inspiration
from Postnikov \cite{Postnikov}, one finds that for a given number of particles $n$ and
given helicity $k$ one should take the lattice in Fig.~\ref{fig.lattice}

\noindent and translate it with the use of the dictionary of Fig.~\ref{fig.dictionary}

\noindent into trivalent ``plabic'' diagrams, which in this case are planar diagrams with only three-point white and
black vertices. Then one identifies all black vertices with
$\mathcal{R}_{\bullet}$ and all white vertices with $\mathcal{R}_\circ$.
The formula for the tree-level harmonic R-matrix $\mathcal{R}_{n,k}$ is
obtained by multiplying all three-point R-matrices appearing in the
plabic diagram and integrating over internal, on-shell propagators, which
reduces to solving a set of linear equations. 
In generalization of our previous analysis we also assign a non-vanishing central charge to external particles. 
After a systematic study
of this gluing procedure one realizes that the final formula for $\mathcal{R}_{n,k}$
depends on $k \,(n-k)$ spectral parameters which can be identified with
the number of faces in the lattice in Fig.~\ref{fig.lattice}. To be more specific, the spectral-parameter dependence appears in the form $\prod_i f_i^{-1+z_i}$ 
in the integrand,
where $f_i$ are the face variables of the plabic diagram, $z_i$ are any complex numbers, and the product is taken over all faces. 
In our interpretation the spectral parameters  $z_i$  correspond to the ``unquantized'' helicities of the particles circling the loops of the plabic diagrams.

\section{Loop amplitudes and spectral regularization}
\label{sec.specreg}

In the following preliminary study we restrict ourselves to the simplest case of the one-loop four-point amplitude. Without deformation,  the computation for $\NN = 4$ SYM
results in the factorization of the tree-level amplitude times the scalar box integral 
\begin{equation}
\label{oneloop}
\mathcal{A}_{4,2}^{\mbox{\tiny 1-loop}}=\mathcal{A}_{4,2}^{\mbox{\tiny tree}}\int d^4 q \frac{(p_1+p_2)^2(p_1+p_4)^2}{q^2 (q+p_1)^2 (q+p_1+p_2)^2 (q-p_4)^2} \,.
\end{equation}
The integration over the loop momentum leads to infrared divergences and thus requires regularization. The most common procedure is dimensional 
regularization see however
\footnote{An alternative massive regularization respecting dual conformal symmetry was proposed in L.~F.~Alday, J.~M.~Henn, J.~Plefka and T.~Schuster,
JHEP {\bf 1001} (2010) 077,
it is not related to our proposal.}. We will avoid it here.

\begin{figure}[h]
\psset{unit=0.8cm}
\begin{pspicture}(5,4.8)
\rput(1,1){\rnode{A1}{}}
\rput(3,1){\rnode{A2}{}}
\rput(4,1){\rnode{A3}{}}
\rput(1,2){\rnode{A4}{}}
\rput(2,2){\rnode{A5}{}}
\rput(3,2){\rnode{A6}{}}
\rput(1,3){\rnode{A7}{}}
\rput(2,3){\rnode{A8}{}}
\rput(3,3){\rnode{A9}{}}
\rput(1,4){\rnode{A10}{}}
\rput(3,4){\rnode{A11}{}}
\rput(4,4){\rnode{A12}{}}
\ncline{A1}{A3}
\ncline{A1}{A10}
\ncline{A10}{A12}
\ncline{A12}{A3}
\ncline{A11}{A2}
\ncline{A7}{A9}
\ncline{A4}{A6}
\ncline{A5}{A8}
\rput(0.5,0.5){\rnode{B1}{3 }}
\rput(4.5,0.5){\rnode{B2}{ 4}}
\rput(0.5,4.5){\rnode{B3}{2 }}
\rput(4.5,4.5){\rnode{B4}{ 1}}
\ncline{A1}{B1}
\ncline{A3}{B2}
\ncline{A10}{B3}
\ncline{A12}{B4}
\pscircle[fillstyle=solid,fillcolor=white](1,1){0.15}
\pscircle[fillstyle=solid,fillcolor=black](3,1){0.15}
\pscircle[fillstyle=solid,fillcolor=white](4,1){0.15}
\pscircle[fillstyle=solid,fillcolor=white](1,2){0.15}
\pscircle[fillstyle=solid,fillcolor=black](2,2){0.15}
\pscircle[fillstyle=solid,fillcolor=white](3,2){0.15}
\pscircle[fillstyle=solid,fillcolor=black](1,3){0.15}
\pscircle[fillstyle=solid,fillcolor=white](2,3){0.15}
\pscircle[fillstyle=solid,fillcolor=black](3,3){0.15}
\pscircle[fillstyle=solid,fillcolor=black](1,4){0.15}
\pscircle[fillstyle=solid,fillcolor=white](3,4){0.15}
\pscircle[fillstyle=solid,fillcolor=black](4,4){0.15}
\rput(2.5,0.5){\rnode{C1}{$0$}}
\rput(2,1.5){\rnode{C1}{$-2\bar\epsilon$}}
\rput(1.5,2.5){\rnode{C1}{$-3\bar\epsilon$}}
\rput(2.5,2.5){\rnode{C1}{$0$}}
\rput(0.5,2.5){\rnode{C1}{$-4\bar\epsilon$}}
\rput(3.5,2.5){\rnode{C1}{$3\bar\epsilon$}}
\rput(4.5,2.5){\rnode{C1}{$4\bar\epsilon$}}
\rput(2.5,4.5){\rnode{C1}{$0$}}
\rput(2,3.5){\rnode{C1}{$-2\bar\epsilon$}}
\end{pspicture}
\caption{Plabic diagram for the one-loop four-point MHV case. A regulating assignment of spectral parameters is added. Note that the spectral parameters of the external and internal lines are the {\it difference} of the numbers assigned to the faces. For instance, the parameter associated to the line connecting particles 1 and 4 is $z = 4\bar\epsilon - 3 \bar\epsilon = \bar\epsilon$, with the sign being determined by the choice of helicity flowing upward.}
\label{fig.oneloop.plabic}
\end{figure}
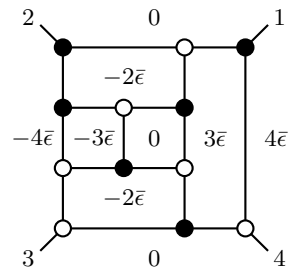

Let us first suppress the $z$-dependence and reproduce the unregulated result in \eqref{oneloop} as proposed in \cite{ArkaniHamed}. We choose the following parametrization of  on-shell momenta
\begin{equation}
\label{parametrization}
p^{\alpha \dot\alpha}= \lambda^{\alpha}\tilde\lambda^{\dot \alpha} = t \left(\begin{tabular}{c}1\\$x$\end{tabular}\right)\,\cdot \left(1\,\,y\right) =\left( \begin{tabular}{cc}$t$&$t \,y$\\$t\,x$&$t\,x\,y$\end{tabular}\right) \,.
\end{equation}
The one-loop four-point MHV amplitude may be obtained from a large number of equivalent plabic diagrams \cite{ArkaniHamed}. We found the diagram in Fig.~\ref{fig.oneloop.plabic} particularly useful for our purposes.
%
The procedure to obtain the box integral is clear from the previous sections: one has to glue three-point MHV and $\overline{\mathrm{MHV}}$ amplitudes as in Fig.~\ref{fig.oneloop.plabic}. Counting the number of delta functions and integrations, one easily sees that four variables are left unintegrated. Further, these are exactly the four integrations which reconstruct the off-shell momentum of the loop integration \cite{ArkaniHamed}
\begin{equation*}
\int \frac{d^4 q}{q^2}=\int \frac{d^2\lambda \,d^2\tilde\lambda}{{\rm GL}(1)}\frac{d \tau}{\tau}=\int t \, dt\, dx\, dy\, \frac{d\tau}{\tau} \,,
\end{equation*}
with the off-shell momentum written in terms of $p^{\alpha\dot\alpha}$, parametrized as in  (\ref{parametrization}), and reference spinors $\lambda^\alpha_1$ and $\tlambda^{\dot\alpha}_4$ associated to, respectively, external particles 1 and 4
\begin{equation*}
q^{\alpha\dot\alpha}=p^{\alpha\dot\alpha}+\tau \lambda^\alpha_4 \tlambda^{\dot\alpha}_1 \,.
\end{equation*}
Up to a trivial numerical factor, this procedure yields the IR-divergent one-loop four-point amplitude (\ref{oneloop}).

We now introduce spectral parameter dependence into the above calculation, replacing the three-point amplitudes by the three-point harmonic   R-matrices  $\mathcal{R}_{\bullet}(z_1,z_2)$ and
$\mathcal{R}_\circ(z_1,z_2)$, {\it cf}~\eqref{Rz3}. 
 A particular, suitable choice of spectral parameters is 
shown in Fig.~\ref{fig.oneloop.plabic},  resulting in the following multiplicative regulating modification of the integrand of the box integral in \eqref{oneloop}
\begin{equation*}
\frac{(\langle 34\rangle[21])^{-4\bar\epsilon}}{q^{-2\bar\epsilon}(q+p_1)^{-2\bar\epsilon}(q+p_1+p_2)^{-2\bar\epsilon}(q-p_4)^{-2\bar\epsilon} }\, .
\end{equation*}
It is reminiscent of analytic regularization, see \cite{Smirnov} and references therein. 
We then see that the spectral parameter can be used in our one-loop example as a regulator, while staying in exactly four dimensions! It should be noted, however, that this choice is not unique and other choices can have a non-regulating effect. We suspect this embarrassment of riches to be solved \emph{via} first principles.

\vspace{-0.2cm}
\section{Conclusions and outlook}
\label{sec.outlook}

In this letter we propose a new way of looking at the interplay between scattering amplitudes and integrability. By solving Yang-Baxter as well as bootstrap equations in the Gra\ss mannian language, we have been able to introduce the notion of {\it spectral parameter} into the scattering problem of $\mathcal{N}=4$ SYM. These parameters have the mathematical interpretation of particle central charges, and the physical interpretation of unquantized, complex helicities. We have presented initial evidence that the deforming parameters may be used to replace dimensional regularization by {\it spectral regularization}. Considering the IR-divergent one-loop scalar-box integral, we have shown that a suitable $z$-deformation indeed regulates the integral. It is important to stress that the regulator is not {\it ad hoc}, but naturally emerges from integrability. 

In conjunction with the crucial insights of \cite{ArkaniHamed}, our results call for a large number of exciting follow-up investigations. The most urgent issue is to establish that IR spectral regularization works to arbitrary loop order, and that it is consistent: E.g.~it needs to be established that the regulator properly exponentiates at higher loop order. This might significantly reduce the deformation freedom, i.e.~might put strong constraints on the set of spectral parameters. In \cite{ArkaniHamed} it is stressed that the general $\mathcal{N}=4$ loop integrand is a differential form with structure $\prod_i d  \log f_i$, where $f_i$ are the face variables mentioned in section \ref{sec.blocks}. Roughly speaking, this should turn into $\prod_i d (\frac{1}{z_i} f_i^{z_i})$ under spectral regularization. If true, this should open the way for a completely new, symmetry respecting technical approach to loop calculations, replacing dim reg. More generally, we suspect that spectral regularization might also be a natural UV regulator, wherever needed (Wilson loops, correlation functions, form factors, etc.). However, the most exciting perspective is to get a handle on all-loop, i.e.~(planar) non-perturbative calculations by applying the powerful techniques of the two-dimensional {\it quantum inverse scattering method} to our four-dimensional system. Recall that in the $\mathcal{N}=4$ spectral problem the one-loop spectral parameter is ``split'' into two parameters $x^\pm$ by the coupling constant \cite{Beisert:2005fw}. Can we further deform our R-matrices to include the coupling in a non-perturbative fashion? Finally, we find it exciting to investigate whether locally ``unquantizing'' the helicities of massless particles could lead to new ways to regulate IR and UV infinities in more general, non-integrable quantum field theories.

\vspace{-0.3cm}
\section{Acknowledgements}

%

We would like to thank N.~Arkani-Hamed, R.~Frassek, J.~Henn, T.~McLoughlin and C.~Sieg for useful discussions. T.~\L .~, J.~P.~and M.~S.~thank the {\it Israel
Institute for Advanced Studies} in Jerusalem for hospitality during the initial stage of this work.
J.~P.~and L.~F.~are supported by the VW-Foundation. C.~M.~is partially supported by a DFG grant of the SFB 676.

\end{document}